\documentclass[12pt, a4paper]{article}
\usepackage{graphicx}
\usepackage{amssymb}
\usepackage{amsmath}
\usepackage{bm}
\usepackage{color}
\usepackage{theorem}
\usepackage{subfigure}
\usepackage{dsfont}
\usepackage{cite}

\usepackage[sort&compress,numbers, merge]{natbib}

\definecolor{Orange}{cmyk}{0,0.61,0.87,0}
\definecolor{JungleGreen}{cmyk}{0.99,0,0.52,0}
\definecolor{OliveGreen}{cmyk}{0.64,0,0.95,0.40}
\definecolor{Brown}{cmyk}{0,0.81,1,0.60}
\definecolor{RoyalBlue}{cmyk}{0.71,0.53,0,0.12}

\usepackage{xcolor} \usepackage{ulem} \usepackage{changebar}

\newcommand{\bX}{{\boldsymbol X}}

\newcommand{\bD}{{\boldsymbol D}}

\newcommand{\bH}{{\boldsymbol H}}

\newcommand{\bW}{{\boldsymbol W}}
\newcommand{\bQ}{{\boldsymbol Q}}

\newcommand{\bT}{{\boldsymbol T}}

\newcommand{\bPhi}{{\boldsymbol \Phi}}
\newcommand{\D}{{\mathcal{D}}}

\DeclareRobustCommand{\Eq}[1]{Eq.~(\ref{#1})}
\DeclareRobustCommand{\Ref}[1]{Ref.~\cite{#1}}
\DeclareRobustCommand{\Refs}[1]{Refs.~\cite{#1}}
\DeclareRobustCommand{\Fig}[1]{Fig.~\ref{#1}}

\newcommand{\Slash}[1]{{\ooalign{\hfil/\hfil\crcr$#1$}}}

\usepackage[colorlinks=true, linkcolor=OliveGreen, citecolor=RoyalBlue,
urlcolor=Brown]{hyperref}

\begin{document}

\begin{titlepage}

\begin{flushright}
UMN--TH--3605/16\\
FTPI--MINN--16/27
\end{flushright}

\vskip 1.35cm
\begin{center}

{\LARGE
{\bf Neutral Naturalness with\\[10pt] Bifundamental Gluinos}
}

\vskip 1.5cm

{\sc Tony Gherghetta}$^a$,
{\sc Minh Nguyen}$^a$,
and
{\sc Zachary Thomas}$^b$

\vskip 0.5cm

{\it\small $^a$School of Physics \& Astronomy, University of Minnesota,\\
 Minneapolis, Minnesota 55455, USA}\\[3pt]
{\it \small $^b$William I. Fine Theoretical Physics Institute, School of
 Physics \& Astronomy, \\ University of Minnesota, Minneapolis, Minnesota 55455,
 USA}

\date{\today}

\vskip 1.5cm

\begin{abstract}
We study constraints on one-loop neutral naturalness at the LHC by considering gluon partners which are required
to ameliorate the tuning in the Higgs mass-squared arising at two loops. This is done with a simple orbifold model 
of folded supersymmetry which not only contains color-neutral stops but also bifundamental gluinos that are 
charged under the Standard Model color group $SU(3)_C$ and a separate $SU(3)_C'$ group. The bifundamental 
gluinos reduce the Higgs mass tuning at two loops and maintain naturalness provided the gluinos are lighter than 
approximately 1.9 TeV for a 5 TeV cutoff scale. Limits from the LHC already forbid bifundamental gluinos below 
1.4 TeV, and other non-colored states such as electroweakinos, $Z'$ bosons and dark sector bound states may be 
probed at future colliders. The search for bifundamental gluinos therefore provides a direct probe of one-loop 
neutral naturalness that can be fully explored at the LHC.

\end{abstract}

\end{center}
\end{titlepage}

\section{Introduction}

The Higgs boson discovery and the conspicuous absence of physics beyond the standard model at the Large Hadron Collider (LHC) 
has exacerbated the tuning in natural solutions of the hierarchy problem. This has led to a renewed interest in exploring alternative
ways to ameliorate this tuning in order to restore naturalness at the TeV scale that is compatible with LHC constraints.
Solutions to the hierarchy problem of the weak scale generically involve adding top partners to cancel the one-loop divergent contributions to the Higgs mass from the top quark.  In the minimal supersymmetric Standard Model (MSSM), for example, these are the stops, scalars with the same quantum numbers as the top quarks.  As has been noted repeatedly in the literature~\cite{Chacko:2005pe, Burdman:2006tz, Cai:2008au}, the partners need not be charged under the Standard Model color group $SU(3)_C$, for the cancellation to proceed; all that is strictly required is that the multiplicity and coupling strength of partners to the Higgs boson be tied to that of the top quarks.  In the example we consider in this paper, folded supersymmetry models~\cite{Burdman:2006tz,Chacko:2008cb} have stops that are triplets under a different $SU(3)$ than that of the Standard Model, giving the appropriate cancellation for the Higgs mass despite being neutral under $SU(3)_C$.

Including such color-neutral stops can resolve the (little) hierarchy problem at one loop, allowing the natural scale of new physics to be pushed up to around the TeV scale.  Expanding such a solution to cancel two-loop quadratic divergences (or one-loop quadratic divergences in powers of couplings that are not the top Yukawa or the $SU(3)_C$ coupling strength, $\alpha_s$) is non-trivial, and will generically yield particles charged under $SU(3)_C$ that can be easily observable at the LHC.  The major such divergences, assuming a cutoff scale $\Lambda$, are:
\begin{itemize}
\item One-loop contributions to the Higgs mass from its electroweak coupling: $\frac{3 \alpha_W}{4 \pi} \Lambda^2$ 
\item One-loop contributions to the stop masses from their coupling to the Higgs: $\frac{y_t^2}{8 \pi^2} \Lambda^2$
\item One-loop contributions to the stop masses from their strong coupling: $\frac{4 \alpha_s}{3 \pi} \Lambda^2$ (and the related two-loop contributions to the Higgs mass proportional to $\alpha_s y_t^2$). 
\end{itemize}
Together, these would yield an approximate fine-tuning at the percent level for $\Lambda = 5$ TeV and a neutral stop mass of $500$ GeV. These divergences therefore suggest that additional particles, beyond the color-neutral stops, should be observed before 5 TeV to ameliorate this fine-tuning.

The first divergence would suggest the addition of particles like winos and Higgsinos, as in the MSSM.  The second would again suggest Higgsinos, but also, unlike the MSSM, additional uncolored top-like fermions (as the neutral stops cannot couple to the Standard Model 
(SM) top quarks without introducing unwanted additional colored states).  The most important divergence for LHC phenomenology, however, is the last one, which would seem to require that the stops are charged under a dark $SU(3)$ (specifically, not that of the Standard Model) with coupling strength $\alpha_s$.  This, in turn, introduces a one-loop divergent contribution for the stop masses proportional to $\alpha_s$, which will need to be cancelled by a contribution from a gluon partner, such as a gluino.  If such a gluino is in the low-energy spectrum and charged under $SU(3)_C$, it will have the most relevance to LHC phenomenology.

\section{A Natural Folded Supersymmetry Model}
\label{sec:model}

A model with the desired particle content and relations amongst coupling constants can be found using an orbifold construction.  We expand the gauge symmetries of the MSSM, then orbifold by a $\mathds{Z}_2$ to reduce our theory to have the appropriate particle content and symmetries.\footnote{We use a $\mathds{Z}_2$ orbifold to obtain a minimal model, though larger orbifolds can yield interesting features, some of which will be noted below.}  Explicitly, we expand the $SU(3)_C \times SU(2)_L \times U(1)_Y$ of the Standard Model to $SU(6)_C \times SU(4)_L \times SU(2)_F \times U(1)_Y$, under which the Higgs and third-generation quark superfields have the following representations:
\begin{align}
\begin{array}{c | c c c c }
& SU(6)_C & SU(4)_L & SU(2)_F & U(1)_Y \\ \hline 
\bH & \boldsymbol 1 & \boldsymbol 4 & \boldsymbol 2 & +1/2 \\
\bT & \overline{\boldsymbol 6} & \boldsymbol 1 & \boldsymbol 2 & - 2/3 \\
\bQ & \boldsymbol 6 & \overline{\boldsymbol 4} & \boldsymbol 1 & + 1/6
\end{array}
\end{align}
This allows the following familiar superpotential term:
\begin{align}
\bW & = y_t {\rm Tr}[\bH \bT \bQ].
\end{align}
This term is functionally the same mother theory as in the simplest orbifold Higgs model in \Ref{Craig:2014aea}, but with added supersymmetry.  The ``flavor" symmetry, $SU(2)_F$, has no relation to the generations of the Standard Model; it is necessary to ensure that enough particles remain to remove quadratic divergences associated with the Yukawa coupling $y_t$ after orbifolding. Note that this mother theory has hypercharge anomalies as written, but it will ultimately be non-anomalous after the orbifolding.  

We now orbifold by the discrete symmetry
\begin{align}
 \mathds{Z}_2^O = \mathds{Z}_2^{C} \times \mathds{Z}_2^L \times \mathds{Z}_2^F \times \mathds{Z}_2^R.
\end{align}
  The first three symmetries act with $+1 (-1)$ on the top (bottom) halves of fundamentals of the corresponding continuous symmetry in the mother theory, while $\mathds{Z}_2^R$ is the conventional $R$-parity of the MSSM.  This orbifolding breaks the symmetries of the mother theory as follows:
\begin{eqnarray}
& &\left(SU(6)_C \times SU(4)_L \times SU(2)_F \times U(1)_Y\right) / \mathds{Z}_2^O   \nonumber \\
& & \rightarrow SU(3)_C \times SU(3)'_C \times U(1)_C \times SU(2)_L \times SU(2)'_L \times U(1)_L \times U(1)_F \times U(1)_Y.
\end{eqnarray}
The gauge groups $SU(3)_C$ and $SU(2)_L$ are the non-abelian gauge symmetries of the Standard Model, while their primed equivalents are new ``dark" gauge symmetries.  Note that the orbifolding also produces two $U(1)$ gauge symmetries, $U(1)_C$ and $U(1)_L$, related to the corresponding non-abelian gauge symmetries; we expect these to be broken at low energy.  A residual $U(1)_F$ flavor symmetry remains at the classical level, which is equivalent to a linear combination of $B-L$ and 
hypercharge.\footnote{Furthermore, as in the simplest orbifold Higgs model~\cite{Craig:2014roa}, there is an accidental $SU(4)$ global symmetry. However this symmetry will be explicitly broken by supersymmetry-breaking mass terms, since in our setup the Higgs boson will be protected by supersymmetry and not a global symmetry.}

The structure of the orbifold may be better understood by looking at the components of the mother-theory superfields afterwards.  The components that survive the orbifolding are those that are unaffected when acted upon by $\mathds{Z}_2^O$:
\begin{align}
\bW_C & =  \left\{  \left( \begin{array}{c c} g_{A A} & 0 \\ 0 & g_{B B} \end{array} \right) + \frac{Z'_C}{2 \sqrt{3}} \left(\begin{array}{c c} \mathds{1} & 0 \\ 0 & - \mathds{1} \end{array}\right),  \left( \begin{array}{c c} 0 & \widetilde{g}_{A B} \\ \widetilde{g}_{B A} & 0 \end{array} \right) \right\}, \\
\bW_L & =   \left\{  \left( \begin{array}{c c} W_{a a} & 0 \\ 0 & W_{b b} \end{array} \right) + \frac{Z'_L}{2 \sqrt{2}} \left(\begin{array}{c c} \mathds{1} & 0 \\ 0 & - \mathds{1} \end{array}\right) ,  \left( \begin{array}{c c} 0 & \widetilde{W}_{a b} \\ \widetilde{W}_{b a} & 0 \end{array} \right) \right\}, \\
\bQ & =   \left\{  \left( \begin{array}{c c} 0 & \widetilde{q}_{A b} \\ \widetilde{q}_{B a} & 0 \end{array} \right),  \left( \begin{array}{c c} q_{A a} & 0 \\ 0 & q_{B b} \end{array} \right) \right\}, \\
\bT & =   \left\{  \left( \begin{array}{c c} 0 & \widetilde{t}_{\alpha B} \\ \widetilde{t}_{\beta A} & 0 \end{array} \right),  \left( \begin{array}{c c} t_{\alpha A} & 0 \\ 0 & t_{\beta B} \end{array} \right) \right\}, \\
\bH & =   \left\{  \left( \begin{array}{c c} H_{a\alpha} & 0 \\ 0 & H_{b \beta} \end{array} \right),  \left( \begin{array}{c c} 0 & \widetilde{H}_{a \beta} \\ \widetilde{H}_{b \alpha} & 0 \end{array} \right) \right\},
\end{align}
with $A$, $a$, $\alpha$ representing color, electroweak, and $SU(2)_F$ ``flavor" indices respectively; we will suppress the latter going forward.  The $U(1)_C$ and $U(1)_L$ gauge bosons are denoted as $Z'_C$ and $Z'_L$, respectively; they effectively couple with half the strength of a corresponding non-abelian boson, as a result of the $\mathds{Z}_2$ orbifold. Note that the surviving gluinos, $\widetilde g$, and winos, $\widetilde W$, are bifundamentals under $SU(3)_C\times SU(3)'_C$, and $SU(2)_L\times SU(2)'_L$, respectively.

With these particles alone, the gauge groups are anomalous.  We fix this mostly by adding additional fermions: we cancel some anomalies in a chiral fashion by adding the rest of the SM third generation fermions (and the complete first two generations as well), and other anomalies in a vector-like fashion by adding Dirac partners for the $B$-tops and the Higgsinos.  We do not need or expect the corresponding scalars of the new fermions to be present; even the second Higgs doublet, $H_d$, is unnecessary, as lepton and down-type quark masses can arise from hard supersymmetry-breaking terms like $H^* q d$ (or their superspace equivalent $[\bX^\dagger \bH^\dagger \bQ \bD]_D$, with $\bX$ a supersymmetry-breaking spurion).

The diagonal abelian gauge symmetries, $U(1)_C$ and $U(1)_L$, are still anomalous after such additions; this is not a problem as long as we take the corresponding gauge bosons to be massive, as is desired on phenomenological grounds anyway.

\subsection{Contributions to scalar masses}
To see how this construction protects our scalar masses, let us explicitly write down the resulting quartic and Yukawa couplings proportional to $y_t$ (leaving out $\!H_{b}$ and the $A$-stops $\widetilde{q}_{A b}, \widetilde{t}_A$, which are unimportant for one-loop naturalness):
\begin{align}
\mathcal{L} & = - y_t H_{a} q_{A,a} t_A - y_t \widetilde{t}_B q_{B,b} \widetilde{H}_{b} - y_t \widetilde{q}_{B,a} t_B \widetilde{H}_{a} + {\rm h.c.}\nonumber \\ 
& \quad \, - | y_t H_{a} \widetilde{q}_{B,a} |^2 - |y_t H_{a} \widetilde{t}_B |^2 - |y_t \widetilde{q}_{Ba} \widetilde{t}_B|^2~.
\label{eq:ytterms}
\end{align}
Note that all the scalars, in particular the $B$-stops, couple to fermions, thereby ensuring that one-loop quadratic divergences to the masses cancel. This is a consequence of the SU(2)$_F$ ``flavor" symmetry and differs from the original folded supersymmetry~\cite{Burdman:2006tz}.

\subsubsection{Higgs}
The Higgs is protected from quadratic divergences by $B$-stop loops.  Dropping all dependence on electroweak symmetry breaking, we have\footnote{We neglect differences between the different $y_t$ in \Eq{eq:ytterms} due to RG running, which can reintroduce a small quadratic divergence.}
\begin{align}
\Delta m_H^2 & = - \frac{3}{8 \pi^2} y_t^2 m_{\widetilde{t}}^2 \log \tfrac{\Lambda^2}{m_{\widetilde{t}}^2} - \frac{3 |A_t|^2}{16 \pi^2} \left(\log \tfrac{\Lambda^2}{m_{\widetilde{t}}^2} - 1 \right) 
\label{eq:stopcontributions}
\end{align}
The first term tells us, as usual, that the $B$-stops cannot be too heavy.  If we demand that contributions to the Higgs mass-squared be no larger than $(200 \textrm{ GeV})^2$ (i.e. the tuning should be no more than 1 part in 5), this puts an upper limit of roughly $470$ GeV on the stop mass for a $\Lambda = 5$ TeV benchmark.  The second term arises from $A$-terms between the Higgs and the $B$-stops;\footnote{For a larger orbifold, such $A$-terms may not exist, due to the mother theory's discrete $R$-symmetry.} naturalness demands that they cannot be too large, but they can still be large enough to yield a physical Higgs mass of $125$ GeV. The physical Higgs mass is (taking, for simplicity, a common stop mass) 
\begin{align}
m_h^2 & = M_Z^2 + \tfrac{1}{2} M_W^2 +  \frac{3 m_t^4}{2 \pi^2 v^2} \left( \log \tfrac{m_{\widetilde{t}}^2}{m_t^2} + \frac{|A_t|^2}{m_{\widetilde{t}}^2} - \frac{|A_t|^4}{12 m_{\widetilde{t}}^4} \right) .
\end{align}
The first term is the usual tree-level expectation; there is no factor of $\cos^2 2 \beta$ here as we are working in a one-Higgs doublet model.  The second contribution arises from the extra $U(1)_L$ $D$-term, which boosts the tree-level mass to $107$ GeV. In fact, even in the absence of $A$-terms, a stop mass of 450 GeV is enough to yield a $125$ GeV Higgs.  If in our mother theory we had gauged $U(4)$ rather than $SU(4)$, we would have $M_Z^2 + M_W^2 \approx $ ($121$ GeV)$^2$ at tree level instead, and one might have to worry about having the loop-level terms becoming too large.\footnote{A larger orbifold would raise this to $M_Z^2 + (1-\tfrac{1}{\Gamma}) M_W^2$, with $\Gamma$ the order of the discrete group, again potentially causing problems for large $\Gamma$. Note that this extra tree-level contribution is reminiscent of a 
similar enhancement found in Ref.~\cite{Craig:2013fga}.} The third term arises from the usual top Yukawa radiative correction to the Higgs quartic coupling, except there is no $\mu$ contribution since we are considering a decoupled $H_d$ .

\subsubsection{Stops}
The stops have their $y_t$-associated quadratic divergences removed by loops of Higgsinos and $B$-tops, as can be seen from the terms in \Eq{eq:ytterms}.  The Higgsinos can get their masses from an $SU(4)_L$-respecting superpotential term, analogously to the MSSM; this would also give a contribution to the Higgs mass at tree level, so the $a$-Higgsino would be around the weak scale.  Alternatively, the dominant contribution to their masses can come from supersymmetry-breaking (e.g. from $[\bX^\dagger \bX \D^\alpha \bH_u \D_\alpha \bH_d]_D$), in which case they can be closer to the TeV scale.  The same argument applies for vector-like $B$-top masses.  The resulting contribution to the stop masses (taking, for simplicity, a common mass $m_{1/2}$ for the $B$-tops and Higgsinos)
\begin{align}
\Delta m^2_{\widetilde{t}_B} & = \frac{y_t^2}{2 \pi^2} m_{1/2}^2 \left( \log \tfrac{\Lambda^2}{m_{1/2}^2} - \tfrac{1}{2}\right), \\
\Delta m^2_{\widetilde{q}_B} & = \frac{y_t^2}{4 \pi^2} m_{1/2}^2 \left( \log \tfrac{\Lambda^2}{m_{1/2}^2} - \tfrac{1}{2}\right).
\end{align}
We have neglected here contributions proportional to the Higgs mass, $A$-terms, and the stop masses themselves (including any superpotential contribution to their masses, as we assume tree-level naturalness).  This tells us that the $b$-Higgsinos and the left $B$-tops should be below about $1.5$ TeV; the $a$-Higgsinos and the right $B$-tops are less constrained and can be multi-TeV.  Such heavy, uncolored particles will of course be difficult to probe at the LHC.

More experimentally promising, however, are the naturalness constraints arising from quadratically-divergent contributions to the squark masses associated with $\alpha_s$.  These are cancelled out partially by the bifundamental gluinos, which are, crucially, charged under $SU(3)_C$.  The cancellation is not complete, however, because the mother theory contained $SU(6)$ and not $U(6)$: there are nine bifundamental gluinos, eight adjoint gluons and $D$-terms, and one half-strength singlet gluon and $D$-term.  The singlet gluon is not necessary for naturalness with a 5 TeV cutoff, and we leave it out of the spectrum; the singlet $D$-term remains, however.
The stop mass contribution becomes
\begin{align}
\Delta m^2_{\widetilde{t}} & = - \frac{7 \alpha_s}{48 \pi} \Lambda^2 + \frac{3 \alpha_s }{2 \pi} m_{\widetilde{g}}^2 \log \tfrac{\Lambda^2}{m_{\widetilde{g}}^2}~.
\end{align}
The first term, while still quadratically divergent, is much reduced from the nonsupersymmetric case (a factor of $4 (N^2 -1)=32$ has been reduced to -7/2), and is not constraining for a 5 TeV cutoff.   The second, logarithmic contribution tells us that the bifundamental gluino should be below about $1.2$ TeV for $\Delta m^2_{\widetilde{t}} \simeq m^2_{\widetilde{t}}=$ (470 GeV)$^2$.

The bifundamental gluino has a Dirac mass, but it should be stressed that it arises from a Majorana mass term in the mother theory, $[\bX \bW \bW]_F$, which is not automatically ``supersoft"; the additional chiral multiplet (with its adjoint scalars) required to replicate the $\mathcal{N}=2$ supersoft structure is not present.    One can add such a chiral multiplet to make the gluinos supersoft, writing a Dirac mass term in the mother theory of $[\bW' \bW \bPhi]_F$ (with $\bW'$ a $D$-term supersymmetry-breaking spurion, and $\bPhi$ the required adjoint chiral multiplet).\footnote{This is required to give mass to the gluinos for many orbifolds larger than $\mathds{Z}_2$ (such as $\mathds{Z}_3$), in which a Majorana mass term in the mother theory is forbidden by the discrete $R$-symmetry; in the daughter theory, the putative Dirac partners within the gluino multiplet are removed by the orbifold.} The supersoft partners for the gluino would be accompanied by octet $B$-sgluons, octet $A$-sgluons, and a singlet sgluon.  The latter two are not critical for naturalness and can be made heavy.  When including the $B$-sgluons, the resulting finite contribution dominates over the residual logarithmic one:
\begin{align}
\Delta m^2_{\widetilde{t}} & = - \frac{7 \alpha_s}{48 \pi} \Lambda^2 + \frac{\alpha_s }{6 \pi} m_{\widetilde{g}}^2 \log \tfrac{\Lambda^2}{m_{\widetilde{g}}^2} + \frac{4 \alpha_s}{3 \pi} m_{\widetilde{g}}^2 \log 4 ,
\end{align}
and the gluino mass limit can be pushed up to $1.9$ TeV for each gluino.  We have assumed the 
(CP-even) $B$-sgluon mass is twice the Dirac mass~\cite{Fox:2002bu}, and there are neither additional substantial supersymmetry-breaking contributions to the $B$-sgluon mass (although, the massless pseudoscalar in the $B$-sector can receive supersymmetry-breaking contributions), nor splittings between the two gluino Dirac masses.  There are, of course, also two-loop contributions to the Higgs mass squared proportional to 
$\alpha_s y_t^2$.  These will largely give similar constraints to the above, though note that, crucially, they prevent taking $\alpha_{s,B}$ to be substantially smaller than $\alpha_{s,A}$ in an effort to skirt the above constraints.

\subsubsection{Electroweak and other contributions}
There are also quadratic electroweak divergences to the Higgs mass proportional to $\alpha_W$.  These are similar in form to the stop divergences proportional to $\alpha_s$, except that in this case we may want to keep the $U(1)_L$ gauge boson with mass, $m_{Z'}$ in the spectrum.  The resulting contribution to the Higgs soft mass is
\begin{align}
\Delta m^2_H & = - \frac{\alpha_W}{8 \pi} \Lambda^2 - \frac{3 \alpha_W}{32 \pi} m_{Z'}^2 \log \tfrac{\Lambda^2}{m_{Z'}^2} + \frac{\alpha_W}{ \pi}  m_{\widetilde{W}}^2 \log \tfrac{\Lambda^2}{m_{\widetilde{W}}^2}. \label{eq:electroweak}
\end{align}
As for the stop mass, there is still a quadratically-divergent contribution, because the mother theory gauged $SU(4)_L$ rather than $U(4)_L$.\footnote{For a larger orbifold, the residual quadratic divergence is suppressed by $1/\Gamma$, with $\Gamma$ the order of the discrete group; this may 
allow $\Lambda$ to be increased (along with the particle content).}
The size of this divergence has been decreased by a factor of six, however, from $4 (N^2 - 1)=12$ to $-2$.  This residual quadratic divergence puts the fundamental limit on $\Lambda$, which cannot be taken much above $5$ TeV for $\Delta m^2_H = (200$~GeV)$^2$.  The remaining logarithmic divergences also imply that the winos and the $b$-Higgsinos need to be lighter than around 800 GeV, and that the $U(1)_L$ gauge boson should be below the cutoff.  As with the gluino bounds, the fermion bounds can be weakened to $1.5$ TeV or higher by making the winos supersoft.

Divergences arising from hypercharge are not a concern in this model, since $\textrm{Tr } Y$ for the scalars in our theory is zero, due to the absence of $H_d$.  As a result, we do not need to worry about quadratically divergent contributions to the $U(1)_Y$ Fayet-Iliopoulos term.  Other divergences proportional to hypercharge can be safely ignored, and a bino is not needed for naturalness for a $5$ TeV cutoff.  

Note that we have not yet needed $H_b$, $\widetilde{t}_A$, or $\widetilde{q}_{Ab}$ for the naturalness of the weak scale.  One can lift these scalars above the cutoff, though note that doing so yields quadratically divergent contributions to the $U(1)_C$ and $U(1)_L$ Fayet-Iliopoulos terms, which in turn feed into the squark and Higgs masses:
\begin{align}
(\Delta m^2_{\widetilde{q}_{B}})_{FI} & =  \frac{\alpha_s}{16 \pi} \Lambda^2, \\
(\Delta m^2_{H})_{FI} & = - \frac{\alpha_W}{8 \pi} \Lambda^2.
\label{eq:HDterm}
\end{align}
These are not constraining by themselves, though the last contribution \Eq{eq:HDterm} might be, in combination with the other electroweak contributions in \Eq{eq:electroweak}.  One can remove these divergences, if desired, by restoring the $b$-Higgs and the $A$-squarks to the spectrum in the multi-TeV range.

\subsection{Summary}
In summary, for a $5$ TeV cutoff scale, the following new particles are required for naturalness of the 
weak scale:

\begin{itemize}

\item The vector-like pair of gluinos $\widetilde{g}_{AB}$ and $\widetilde{g}_{BA}$, each a bifundamental under SM and dark $SU(3)_C$, which should be below $1.2$ TeV.  Alternatively, one can have two pairs of bifundamental gluinos along with a dark $SU(3)_C$ octet sgluon, each below $1.9$ TeV.
\item The vector-like pair of winos $\widetilde{W}_{ab}$ and $\widetilde{W}_{ba}$, each a bifundamental under SM and dark $SU(2)_L$.  These should be below $800$ GeV.
\item The Higgsinos $\widetilde{H}_b$ and $\widetilde{H}_a$, which are doublets under dark and SM $SU(2)_L$, respectively, each with Dirac masses. The former should be below around $800$ GeV, while the latter can be in the multi-TeV range.
\item The $W$ gauge bosons of dark $SU(2)_L$.  We take dark $SU(2)_L$ not to be spontaneously broken, so they confine into weak glueballs at roughly the $100$ eV range.\footnote{Note that, due to the absence of light quarks in the dark sector, dark $SU(2)_L$ is not broken by chiral symmetry breaking effects.  One could consider the case where dark $SU(2)_L$ is spontaneously broken by a relatively light $H_b$, but this would require additional particle content at low energy to ensure that such an $H_b$ mass is natural and could cause deviations in the $\rho$-parameter.}
\item The gluons of dark color, which confine into glueballs at masses around the $10$ GeV range.
\item The dark stops, $\widetilde{t}_B$ and $\widetilde{q}_B$, at or below $470$ GeV.  These are singlets under SM color, but are charged under the other SM gauge groups.
\item The vector-like dark tops, $q_B$ and $t_B$; these are singlets under SM color, but are charged under the other SM gauge groups.  The former should be below around $1.5$ TeV, while the latter can be in the multi-TeV range.

\item The gauge boson of the diagonal gauge group $U(1)_L$ should also be below the cutoff.  The Higgs and the left-handed third-generation quarks (at least) are charged under it.
\end{itemize}

The most likely candidate for direct production at the LHC is the bifundamental gluino, as it is charged under SM color.  The $U(1)_L$ gauge boson could also be a candidate for a $Z'$-style search; however, if it does not directly couple to first- or second-generation fermions,\footnote{This is not an \emph{ad hoc} assumption; note that the down-type Yukawas are incompatible with an unbroken $U(1)_L$ symmetry without giving a charge to right-handed quarks, suggesting a non-trivial connection to flavor.} the resulting experimental limits will not be constraining on the naturalness of the theory.

\section{Phenomenological Consequences}
\label{sec:LHC}

\begin{figure}
\centering
\includegraphics[width=0.7\textwidth]{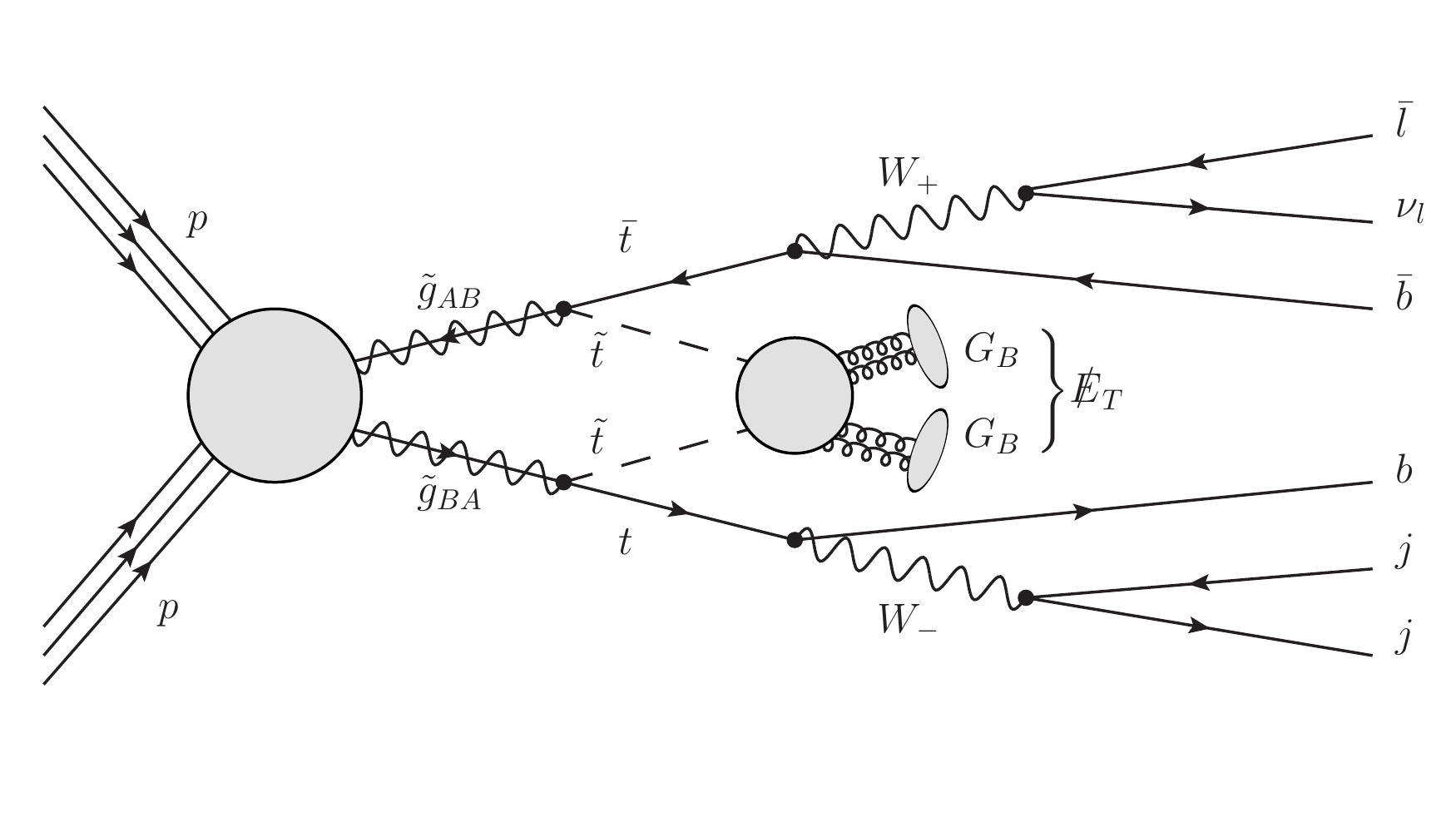}
\caption{A representative Feynman diagram for bifundamental gluino production and decay at a hadron collider.  The stops produced in gluino decay behave as squirks, eventually decaying to dark glueballs, which we take to be invisible on collider scales.}
\label{fig:feynman}
\end{figure}

\subsection{Bifundamental Gluino Searches at the LHC}

The bifundamental gluinos, required in our folded supersymmetry model for two-loop naturalness of the weak scale, are charged under SM color and may thus be easily produced at the LHC.  They will be pair produced in proton-proton collisions, after which they will undergo the decays $\widetilde{g}_{AB} \rightarrow \widetilde{t}_B t_A$ or $\widetilde{g}_{AB} \rightarrow \widetilde{b}_B b_A$.  The lightest squark is individually stable, regardless of whether it is the LSP or not, as it is the lightest (anti-)fundamental under the unbroken dark $SU(3)_C$.  As it is heavier than the confinement scale of dark $SU(3)_C$, the squarks will undergo squirky behavior, radiating off dark glueballs as they are drawn back together to form a non-relativistic bound state, which will then annihilate.  These dark glueballs can have interesting phenomenology, as has been explored elsewhere in the literature~\cite{Kang:2008ea, Juknevich:2009ji, Chacko:2015fbc}; we will assume for our purposes that their lifetime is longer than collider scales, and that any resulting bremsstrahlung of visible particles (mainly expected to be photons) is soft and can be ignored.  As a result, any dark glueballs produced can be treated as missing energy for collider purposes.  

\begin{figure}
\centering
\includegraphics[width=0.5\textwidth]{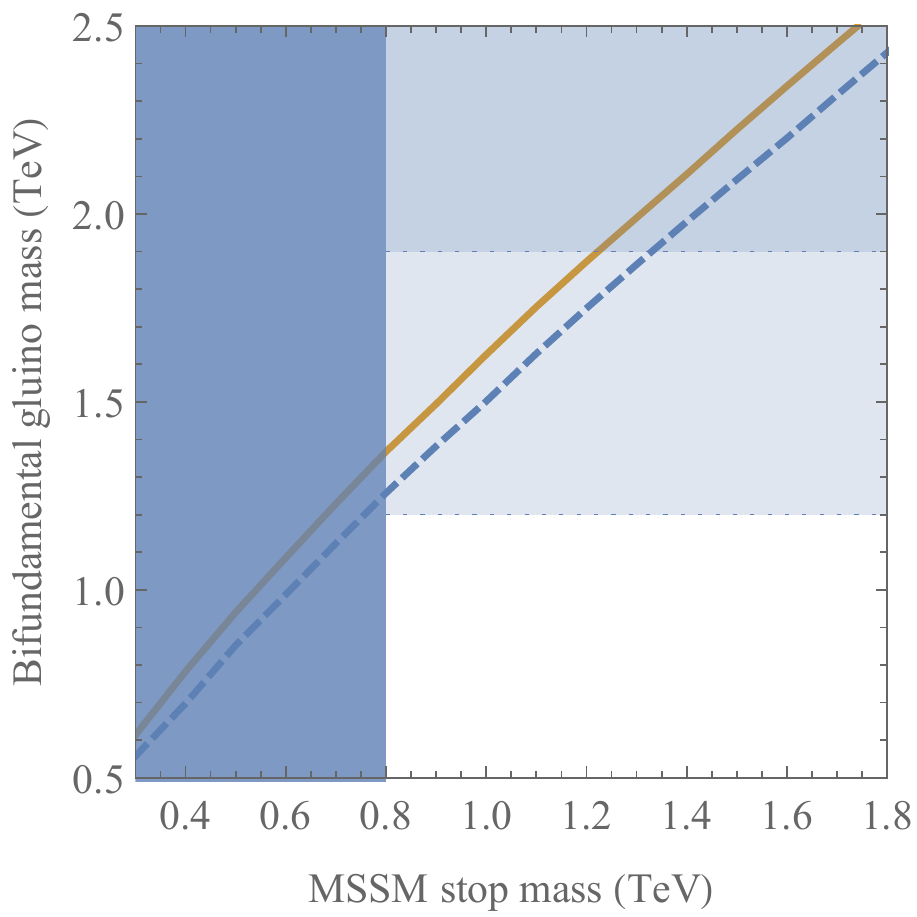}
\caption{Approximate experimental limits for the bifundamental gluino mass in the case where there is one Dirac gluino (dashed, blue line) or two ``supersoft" Dirac gluinos (solid, orange line), as a function of the corresponding limit on the MSSM stop mass.  We have assumed an experimental lower limit on the stop mass, $m_{\widetilde{t}} \gtrsim 800$ GeV. The top shaded regions are those in 
which there is some degree of tuning $(\Delta m_{\tilde t}^2 \gtrsim m_{\tilde t}^2)$ in the $B$-stop mass for one Dirac gluino ($\gtrsim 1.2$ TeV) and  two ``supersoft" Dirac gluinos ($\gtrsim 1.9$ TeV).}
\label{fig:limits}
\end{figure}

Thus, the squark bound state will generically decay into dark glueballs, producing missing energy.  However, if the squark bound state is charged (i.e. $\widetilde{b}_B \widetilde{t}_B$), it will necessarily produce a $W$ in the final decay.  In terms of the final decay products, this is the same as $\widetilde{t}_B \widetilde{t}_B$ production, except with somewhat differing kinematics (as one of the $b W$ pairs will not reconstruct to an on-shell top).

As a result, the generic signals of this process are $pp \rightarrow b \bar{b} W^+ W^- + \Slash{E}_T$ and $pp \rightarrow b \bar{b}+ \Slash{E}_T$.  The latter is disfavored by combinatorics (as there are two light stops and only one light sbottom required by naturalness), so we will consider primarily the former; a representative Feynman diagram for this process is shown in \Fig{fig:feynman}. In terms of the final decay products, this is very similar to the MSSM process $pp \rightarrow \widetilde{t} \widetilde{t} \rightarrow t \bar{t} \chi^0_1 \chi^0_1$, so we can use existing MSSM direct stop production searches to place limits on our model.

CMS results from 12.9 fb${}^{-1}$ of collisions at $\sqrt{s} = 13$ TeV yield a limit on the MSSM stop mass from stop pair production of roughly $800$ GeV \cite{Aaboud:2016lwz, Khachatryan:2016xvy}.  Using Madgraph and Pythia \cite{Alwall:2011uj}, we can compare the cross section for bifundamental gluino pair production in our model to stop pair production in the MSSM.  Doing so yields a corresponding 
limit on the bifundamental gluino mass of $1.3$ TeV (for a single Dirac pair) or $1.4$ TeV (for two ``supersoft" Dirac pairs).  This is only an approximate limit, as we do not take into account any kinematic differences between the two production mechanisms (and their resulting effect after cuts), nor any effects of squirkiness. This likely results in these limits being slightly too stringent, as not all events featuring bifundamental gluino pair production will result in $b \bar{b} W^+ W^- + \Slash{E}_T$.  Prospective future limits on the bifundamental gluino mass (as the MSSM stop mass limit increases with more data) are shown in \Fig{fig:limits}.

In particular, in the above analysis we have neglected an additional possible decay channel of the final non-relativistic squark bound state.  By construction, these third-generation squarks have a large quartic coupling with the Higgs, so instead of decaying invisibly into glueballs, it could decay visibly into $hh$, $W^+ W^-$, or $Z Z$.  Such a decay would certainly be striking.  However, these decays can often be subdominant; for example, for small stop mixing, the $hh$ branching ratio is roughly 10\% of that into dark glueballs~\cite{Drees:1993uw, Martin:2008sv, Batell:2015zla, Chacko:2015fbc}. For larger stop mixing,\footnote{Note that for larger orbifolds in which $A$-terms are forbidden by the discrete $R$-symmetry, stop mixing is also forbidden; the usual mixing through the $\mu$ parameter is also shut off as there is no $H_d$ in this model.} the Higgs branching ratio is enhanced, which could lead to interesting signals; we leave this possibility to future work.

\subsection{Tuning constraints}

The overall tuning in these models is quite modest given current experimental limits.  
We define the tuning by:
\begin{equation}
\frac{2\Delta m_H^2}{m_h^2 } \times \frac{\Delta m_{\widetilde{t}}^2}{m_{\widetilde{t}}^2},
\label{eq:tuning}
\end{equation}
where we sum contributions in quadrature in each case, but never take either fraction to be less than one.  
The first term represents a tuning in the Higgs vacuum expectation value, while the second term represents a tuning in the physical stop mass. 

Taking the lower limit of 1.4 (1.3) TeV for the bifundamental gluino seriously, we estimate that the model with doubly (singly) Dirac gluinos is tuned at around 1 part in 8 (10) level.  In the former case, most of the tuning arises from the electroweak contributions to the Higgs mass. Again, if the mother theory gauged $U(4)_L$ rather than $SU(4)_L$, the tuning would be lessened, at the expense of making the physical Higgs mass of 125 GeV more difficult to achieve without its own fine-tuning.

\begin{figure}
\centering
\includegraphics[width=0.5\textwidth]{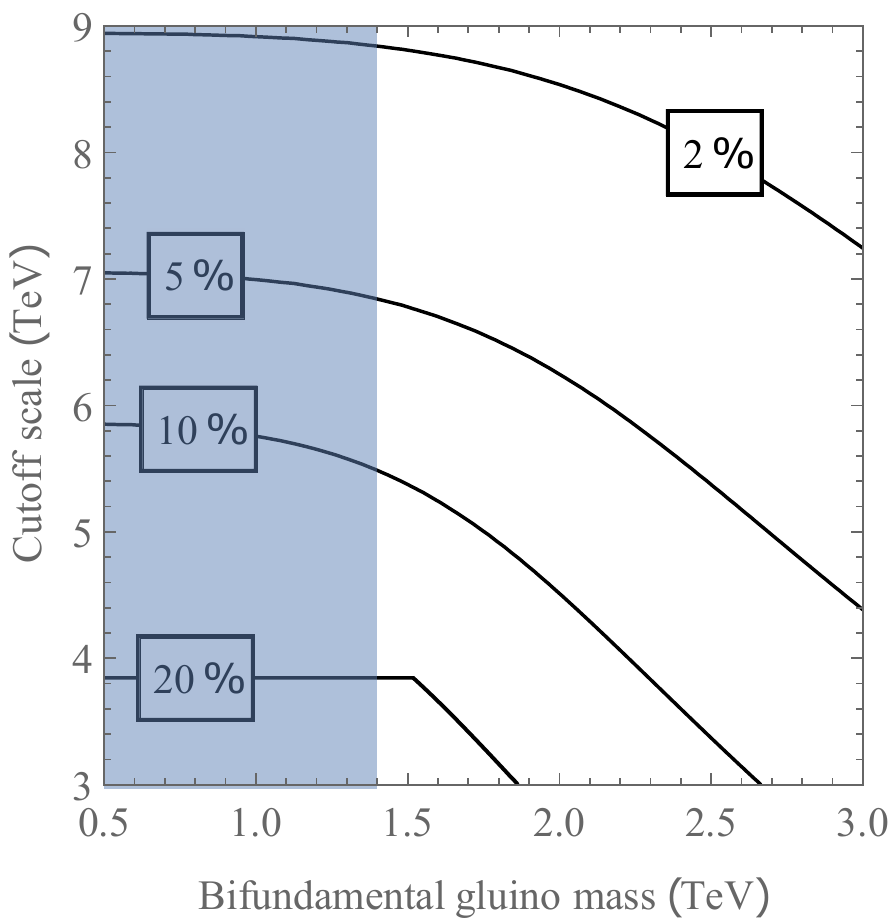}
\caption{Contours of the percent tuning in the model, defined from Eq.(\ref{eq:tuning}), as a function of the cutoff scale $\Lambda$, and the (supersoft) bifundamental gluino mass, $m_{\tilde g}$. The shaded area depicts the excluded “supersoft" Dirac masses below 1.4 TeV  from LHC MSSM stop mass searches. }
\label{fig:tuning}
\end{figure}

A contour plot of the tuning is shown in Figure~\ref{fig:tuning} which shows that, as expected, the tuning is mostly sensitive to an increasing bifundamental gluino mass, since the quadratic divergences are mostly cancelled. This suggests that new colored states should still be accessible at the LHC if naturalness is to remain a relevant criterion.

\section{Conclusion}
In summary, we have considered the addition of colored gluon partners to ameliorate the tuning at two loops in folded supersymmetric models with color-neutral stops. In particular, using a simple $\mathds{Z}_2$ orbifold construction, bifundamental gluinos are found to partly cancel the quadratic divergences. These states are accessible at the LHC, and using current MSSM stop mass limits one can forbid bifundamental gluinos below about 1.3 TeV (1.4 TeV) for a singly (doubly) Dirac pair. Assuming a cutoff scale $\Lambda = 5$~TeV these limits correspond to a natural region of parameter space, and helps to address the little hierarchy problem in supersymmetric models. Other non-colored states, such as electroweakinos, $Z'$ gauge bosons and dark sector bound states remain mostly unconstrained, but may also be accessible in collider experiments. The fact that colored states may be present at two loops suggests that neutral naturalness at one loop could be fully explored at the LHC. Nonetheless, it remains an interesting question whether completely neutral top and gluon partners can be found in a folded supersymmetry model where naturalness is preserved but difficult to probe experimentally.

\section*{Acknowledgments}

We thank Alexsey Cherman, Nathaniel Craig, David Curtin, and Simon Knapen for helpful discussions.
The work of T.G. is supported by the U.S. Department of Energy grant DE-SC0011842 at the
University of Minnesota, and Z.T. is supported by the University of Minnesota.

\section*{Appendix}

\subsection*{Larger Orbifolds}
\renewcommand{\theequation}{A.\arabic{equation}}
\setcounter{equation}{0}

In this paper, we have discussed the case of a $\mathds{Z}_2$ orbifold, but one can easily use a larger discrete group $\mathcal{G}$ of order $\Gamma$ to orbifold, in the manner of \Refs{Craig:2014roa, Schmaltz:1998bg}.  In order to use the power of supersymmetry without having colored stops, the orbifold must include some non-trivial discrete $R$-symmetry.  As this $R$-symmetry must be a subgroup of $U(1)_R$, it must be cyclic.\footnote{There may be less trivial possibilities for extended supersymmetries with larger $R$-symmetries, though incorporating the top Yukawa coupling in a phenomenologically acceptable manner would be difficult.}  This symmetry $\mathds{Z}^R_\rho$ must also be embedded in $\mathcal{G}$, corresponding to one of its one-dimensional representations. 

As an instructive example, let us take the case $\mathcal{G} = \mathds{Z}_\Gamma$ and $\rho = \Gamma$, giving canonical $R$-charge to the superfields; ignoring hypercharge for now, the superfields in the mother theory have the following quantum numbers:
\begin{align}
	\begin{array}{c | c c c c }
		& SU(3 \Gamma)_C & SU(2 \Gamma)_L & SU(\Gamma)_F & \mathds{Z}^R_\Gamma \\ \hline 
		\bH & \boldsymbol 1 & \square & \overline{\square} & 0 \\
		\bT & \overline{\square} & \boldsymbol 1 & \square & +1 \\
		\bQ & \square & \overline{\square} & \boldsymbol 1 & +1 \\
		\bW_{\!\alpha}^C & \textrm{Adj} & \boldsymbol 1 & \boldsymbol 1 & +1 \\
		\bW_{\! \alpha}^L & \boldsymbol{1} & \textrm{Adj} & \boldsymbol 1 & +1 
	\end{array}
\end{align}
Our orbifold reduces the symmetries of the mother theory as follows:
\begin{eqnarray}
& & \left(SU(3 \Gamma)_C \times SU(2 \Gamma)_L \times SU(\Gamma)_F \times \mathds{Z}_\Gamma^R\right) / \mathds{Z}_\Gamma \nonumber \\
& & \rightarrow SU(3)_C^\Gamma \times SU(2)_L^\Gamma \times U(1)_C^{\Gamma-1} \times U(1)_L^{\Gamma -1} \times U(1)_F^{\Gamma -1}
\end{eqnarray}
The surviving non-abelian gauge bosons, quarks, and Higgses are the $\Gamma$ copies along the diagonal, exactly as in \Ref{Craig:2014roa}.  The gauginos, squarks, and Higgsinos, however, have off-diagonal elements due to their non-trivial $R$-charge.  Explicitly:
\begin{align}
	\lambda & = \left(\begin{array}{c c c c c }
		0 & \lambda_{12} & 0 & \cdots & 0 \\
		0 & 0 & \lambda_{23} & \cdots & 0 \\
		0 & 0 & 0 & \cdots & 0 \\
		\cdots & \cdots & \cdots & \cdots & \cdots \\
		\lambda_{\Gamma 1} & 0 & 0 & \cdots & 0 \end{array}\right), \\
	\widetilde{H} & = \left(\begin{array}{c c c c c }
		0 & 0 & 0 & \cdots & \widetilde{H}_{1 \Gamma} \\
		\widetilde{H}_{2 1} & 0 & 0 & \cdots & 0 \\
		0 & \widetilde{H}_{3 2} & 0 & \cdots & 0 \\
		\cdots & \cdots & \cdots & \cdots & \cdots \\
		0 & 0 & 0 & \cdots & 0 \end{array}\right),
\end{align}
with the squarks ($\widetilde{q}_{A b}$, $\widetilde{t}_{\alpha B}$, etc.) following a similar pattern to the gauginos, as they have the same $R$-charge.  Note that for $\Gamma > 2$, the $Z^R_\Gamma$ forbids Majorana gaugino mass terms and $A$-terms in the mother theory; in the daughter theory, there are no fields to build them from.  Dirac gaugino masses may be obtained in the mother theory by introducing adjoint chiral superfields with $R$-charge zero; the surviving fermionic components in the daughter theory follow a similar off-diagonal pattern to the Higgsinos.

The expectations from naturalness for general $\Gamma$ will be similar to the $\Gamma = 2$ case discussed in the paper, except that Dirac-like gaugino masses are mandatory, and additional uncolored squarks will be required to keep the first set of squarks one-loop natural.

One can, of course, use groups where $\mathcal{G}$ is larger than the discrete $R$-symmetry, including non-abelian $\mathcal{G}$.  However, it should be noted that standard model fields necessarily live in one-dimensional irreducible representations of $\mathcal{G}$, and that the off-diagonal supersymmetric particles will only induce couplings between irreducible representations of the same dimension (as the matrices inducing the group action are constructed out of irreps).   Ultimately, one obtains multiple sectors, each of which looks like the cyclic case above, although the $R$-symmetries and irrep dimensions of each may be different.  Some sectors with non-Abelian irreducible representations may even look supersymmetric.  
These different sectors do not couple to each other except through the diagonal $U(1)$ factors and their corresponding $D$-terms (and through hypercharge).

Even if the sector containing the Standard Model looks like the $\mathds{Z}_2$ case, the additional diagonal $U(1)$ factors arising from the $\mathds{Z}_2$ being embedded in some larger group $\mathcal{G}$ of order $\Gamma$ can have definite phenomenological consequences. The corresponding $D$-terms raise the tree-level mass of the SM Higgs, which is now:
\begin{align}
	m_{h,\textrm{tree}}^2 & = M_Z^2 + \left(1 - \tfrac{1}{\Gamma}\right)M_W^2.
\end{align}
The additional $U(1)$s also help to reduce the mismatch between the bosonic and fermionic degrees of freedom. As has been noted in \Ref{Burdman:2006tz}, a $U(\Gamma N)$ parent theory generally gives rise to exact cancellation in one-loop quadratic divergences, while $SU(\Gamma N)$ leaves partial corrections suppressed by $1/(\Gamma N)$.  Increasing $\Gamma$ thereby allows the cutoff to be raised; as $\Gamma$ grows larger, $SU(\Gamma N)$ and $U(\Gamma N)$ become more similar.

\bibliographystyle{JHEP}
\bibliography{ref}

\providecommand{\href}[2]{#2}\begingroup\raggedright\begin{thebibliography}{10}

\bibitem{Chacko:2005pe}
Z.~Chacko, H.-S. Goh, and R.~Harnik, {\it {The Twin Higgs: Natural electroweak
  breaking from mirror symmetry}},  {\em Phys. Rev. Lett.} {\bf 96} (2006)
  231802, [\href{http://arxiv.org/abs/hep-ph/0506256}{{\tt hep-ph/0506256}}].

\bibitem{Burdman:2006tz}
G.~Burdman, Z.~Chacko, H.-S. Goh, and R.~Harnik, {\it {Folded supersymmetry and
  the LEP paradox}},  {\em JHEP} {\bf 02} (2007) 009,
  [\href{http://arxiv.org/abs/hep-ph/0609152}{{\tt hep-ph/0609152}}].

\bibitem{Cai:2008au}
H.~Cai, H.-C. Cheng, and J.~Terning, {\it {A Quirky Little Higgs Model}},  {\em
  JHEP} {\bf 05} (2009) 045, [\href{http://arxiv.org/abs/0812.0843}{{\tt
  arXiv:0812.0843}}].

\bibitem{Chacko:2008cb}
Z.~Chacko, C.~A. Krenke, and T.~Okui, {\it {Supersymmetry in Slow Motion}},
  {\em JHEP} {\bf 01} (2009) 050, [\href{http://arxiv.org/abs/0809.3820}{{\tt
  arXiv:0809.3820}}].

\bibitem{Craig:2014aea}
N.~Craig, S.~Knapen, and P.~Longhi, {\it {Neutral Naturalness from Orbifold
  Higgs Models}},  {\em Phys. Rev. Lett.} {\bf 114} (2015), no.~6 061803,
  [\href{http://arxiv.org/abs/1410.6808}{{\tt arXiv:1410.6808}}].

\bibitem{Craig:2014roa}
N.~Craig, S.~Knapen, and P.~Longhi, {\it {The Orbifold Higgs}},  {\em JHEP}
  {\bf 03} (2015) 106, [\href{http://arxiv.org/abs/1411.7393}{{\tt
  arXiv:1411.7393}}].

\bibitem{Craig:2013fga}
N.~Craig and K.~Howe, {\it {Doubling down on naturalness with a supersymmetric
  twin Higgs}},  {\em JHEP} {\bf 03} (2014) 140,
  [\href{http://arxiv.org/abs/1312.1341}{{\tt arXiv:1312.1341}}].

\bibitem{Fox:2002bu}
P.~J. Fox, A.~E. Nelson, and N.~Weiner, {\it {Dirac gaugino masses and
  supersoft supersymmetry breaking}},  {\em JHEP} {\bf 08} (2002) 035,
  [\href{http://arxiv.org/abs/hep-ph/0206096}{{\tt hep-ph/0206096}}].

\bibitem{Kang:2008ea}
J.~Kang and M.~A. Luty, {\it {Macroscopic Strings and 'Quirks' at Colliders}},
  {\em JHEP} {\bf 11} (2009) 065, [\href{http://arxiv.org/abs/0805.4642}{{\tt
  arXiv:0805.4642}}].

\bibitem{Juknevich:2009ji}
J.~E. Juknevich, D.~Melnikov, and M.~J. Strassler, {\it {A Pure-Glue Hidden
  Valley I. States and Decays}},  {\em JHEP} {\bf 07} (2009) 055,
  [\href{http://arxiv.org/abs/0903.0883}{{\tt arXiv:0903.0883}}].

\bibitem{Chacko:2015fbc}
Z.~Chacko, D.~Curtin, and C.~B. Verhaaren, {\it {A Quirky Probe of Neutral
  Naturalness}},  {\em Phys. Rev.} {\bf D94} (2016), no.~1 011504,
  [\href{http://arxiv.org/abs/1512.05782}{{\tt arXiv:1512.05782}}].

\bibitem{Aaboud:2016lwz}
{\bf ATLAS} Collaboration, M.~Aaboud et~al., {\it {Search for top squarks in
  final states with one isolated lepton, jets, and missing transverse momentum
  in $\sqrt{s}=13$ TeV $pp$ collisions with the ATLAS detector}},  {\em Phys.
  Rev.} {\bf D94} (2016), no.~5 052009,
  [\href{http://arxiv.org/abs/1606.03903}{{\tt arXiv:1606.03903}}].

\bibitem{Khachatryan:2016xvy}
{\bf CMS} Collaboration, V.~Khachatryan et~al., {\it {Search for new physics
  with the MT2 variable in all-jets final states produced in pp collisions at
  sqrt(s) = 13 TeV}},  {\em Submitted to: JHEP} (2016)
  [\href{http://arxiv.org/abs/1603.04053}{{\tt arXiv:1603.04053}}].

\bibitem{Alwall:2011uj}
J.~Alwall, M.~Herquet, F.~Maltoni, O.~Mattelaer, and T.~Stelzer, {\it {MadGraph
  5 : Going Beyond}},  {\em JHEP} {\bf 06} (2011) 128,
  [\href{http://arxiv.org/abs/1106.0522}{{\tt arXiv:1106.0522}}].

\bibitem{Drees:1993uw}
M.~Drees and M.~M. Nojiri, {\it {Production and decay of scalar stoponium bound
  states}},  {\em Phys. Rev.} {\bf D49} (1994) 4595--4616,
  [\href{http://arxiv.org/abs/hep-ph/9312213}{{\tt hep-ph/9312213}}].

\bibitem{Martin:2008sv}
S.~P. Martin, {\it {Diphoton decays of stoponium at the Large Hadron
  Collider}},  {\em Phys. Rev.} {\bf D77} (2008) 075002,
  [\href{http://arxiv.org/abs/0801.0237}{{\tt arXiv:0801.0237}}].

\bibitem{Batell:2015zla}
B.~Batell and S.~Jung, {\it {Probing Light Stops with Stoponium}},  {\em JHEP}
  {\bf 07} (2015) 061, [\href{http://arxiv.org/abs/1504.01740}{{\tt
  arXiv:1504.01740}}].

\bibitem{Schmaltz:1998bg}
M.~Schmaltz, {\it {Duality of nonsupersymmetric large N gauge theories}},  {\em
  Phys. Rev.} {\bf D59} (1999) 105018,
  [\href{http://arxiv.org/abs/hep-th/9805218}{{\tt hep-th/9805218}}].

\end{thebibliography}\endgroup

\end{document}